\documentclass[aip,graphicx]{revtex4-1}

\usepackage{graphicx}
\usepackage{amsmath}

\draft

\begin{document}

\title{Tunable Electronic Interactions and Weak Antilocalization in Bulk Ge$_2$Sb$_2$Te$_{5-5x}$Se$_{5x}$ Phase Change Materials} 

\author{Nicholas Mazzucca}

\affiliation{The Ohio State University, Department of Physics, Columbus, OH, USA}
\author{Junjing Zhao}
\affiliation{University of Virginia, Department of Physics, Charlottesville, VA, USA}
\author{Zhenyang Xu}
\affiliation{University of Virginia, Department of Physics,
Charlottesville, VA, USA}
\author{Despina Louca}
\affiliation{University of Virginia, Department of Physics, Charlottesville, VA, USA}
\author{Utpal Chatterjee}
\affiliation{University of Virginia, Department of Physics, Charlottesville, VA, USA}
\author{Marc Bockrath}
\affiliation{The Ohio State University, Department of Physics, Columbus, OH, USA}

\date{\today}

\begin{abstract}
Phase change materials (PCMs) are well-known for their reversible and rapid switching between crystalline and amorphous phases through thermal excitations mediated by strong electrical or laser pulses. This crystal-to-amorphous transition is accompanied by a remarkable contrast in optical and electronic properties, making PCMs useful in nonvolatile data storage applications. Here, we combine electrical transport and angle resolved photoemission spectroscopy (ARPES) measurements to study the electronic structure of bulk Ge$_2$Sb$_2$Te$_{5-5x}$Se$_{5x}$ (GSST) for $0\le x \le 0.8$, where $x$ represents the amount of Se substituting Te in Ge$_2$Sb$_2$Te$_5$ (GST)-- a prototypical PCM. The single-particle density of states (SDOS) derived from the integrated ARPES data displays metallic behavior for all $x$, as evidenced by the presence of a finite density of states in the vicinity of the chemical potential. Transport measurements also display clear signatures of metallic transport, consistent with the SDOS data. The temperature dependence of the resistance indicates the onset of moderate electron-electron Coulomb interaction effects at low temperatures for $x\geq 0.6$. At the same time, the magnetoresistance data shows signatures of weak antilocalization for $x\geq 0.6$. An analysis on the temperature dependence of the phase coherence length suggests that electron dephasing is primarily due to inelastic electron-electron scattering. We find that these effects are enhanced with increasing $x$, portraying GSST as a novel PCM where electronic interactions can be tuned via chemical doping.
\end{abstract}


\maketitle 

With the ever-expanding adoption of deep learning neural networks, artificial intelligence, machine learning, and large language models to our daily life, there is an almost exponential growth of data being generated in modern times. Efficient storage and convenient access to large amounts of data therefore constitutes a major challenge for contemporary information technology. Long-term solutions will require not only innovative device engineering, but also materials innovation seeking alternatives to silicon-based memory devices, which are close to their fabrication limits. Ge$_2$Sb$_2$Te$_5$ (GST), with existing and matured applications in the device industry, is considered a promising material platform for new-generation memory technology \cite{meinders_optical_2006,yamada_high_1987,ovshinsky_reversible_1968,chen_compound_1986,wuttig_phase-change_2007,kolobov_understanding_2004,sun_structure_2006,lencer_map_2008}. In this context, the large difference in optical conductivity between the crystalline and amorphous phases of phase change materials (PCMs) such as GST has  found full-fledged commercial utilization in digital data storage devices like DVDs and Blu-ray  discs \cite{yamada_phase-change_1998,ohta_phase-change_2001,kuwahara_experimental_2003}. The contrast in electrical resistivity of the PCMs across the crystalline-to-amorphous transition (CAT) is also being actively explored for  phase change random access memory (PCRAM) devices \cite{meng_electrical_2023,nakayama_submicron_2000,sarwat_materials_2017}, which can potentially go far beyond the silicon-based flash state memory technology. Even though GST possesses good thermal stability, excellent scalability, low power consumption, good data retention, and reasonable crystallization speed \cite{guo_review_2019,yamada_rapidphase_1991}, significant optimization of these attributes is needed to enable commercially sustainable GST-based neuromorphic devices. To this end, there has been extensive investigations on the functionalities of doped GST. Recent studies \cite{vinod_direct_2014,vinod_structural_2015,xu_metal-insulator_2020,xu_octahedral_2023} suggest that Ge$_2$Sb$_2$Te$_{5-5x}$Se$_{5x}$ (GSST), i.e., Se doped GST, could potentially be an ideal candidate material due to its (i) significantly swifter switching between crystalline and amorphous phases, (ii) order of magnitude enhancement in the resistivity change across the CAT, (iii) lower power requirement for switching, and (iv) relatively lower toxicity due to reduction in Te content. Given these benefits, there is a strong interest in characterizing the electronic and transport properties of GSST.

Many of the investigations done on GST and GSST have involved only thin film samples. Bulk samples of these materials have not been actively pursued for solid state memory applications due to the perception that they are unable to undergo a CAT. It has, however, recently been established that a CAT can be induced in bulk GSST. Upon increasing $x$, the liquid nitrogen quenched GSST undergoes a CAT as a function of $x$ for $x>0.8$ \cite{xu_metal-insulator_2020}. Even though the crystalline phase of bulk GSST samples are mostly hexagonal, a rocksalt-like structure also coexists as a secondary phase with a volume fraction essentially independent of $x$. Like in thin film samples, the CAT in bulk GSST samples accompanies a metal-to-insulator transition (MIT), with several orders of magnitude increase in the resistivity upon approaching the amorphous state.
Ref.~\citenum{xu_metal-insulator_2020} has also demonstrated the possibility of stabilizing and tuning multiple resistance states by changing $x$ in bulk GSST samples on the metallic side, which can have implications towards the realization of multi-level data storage devices \cite{mehonic_roadmap_2024,boybat_neuromorphic_2018}. To tap into its full application potential, a microscopic picture of carrier propagation in GSST samples is necessary, and this requires a combination of transport and angle-resolved photoemission spectroscopy (ARPES) studies. Although such studies have been conducted in two-dimensional (2D) GST thin films \cite{breznay_weak_2012, kellner_mapping_2018,kim_electronic_2007}, a similar work on bulk three-dimensional (3D) GST or GSST is unavailable, which motivates the present work utilizing ARPES and transport probes.

\begin{figure}
\includegraphics{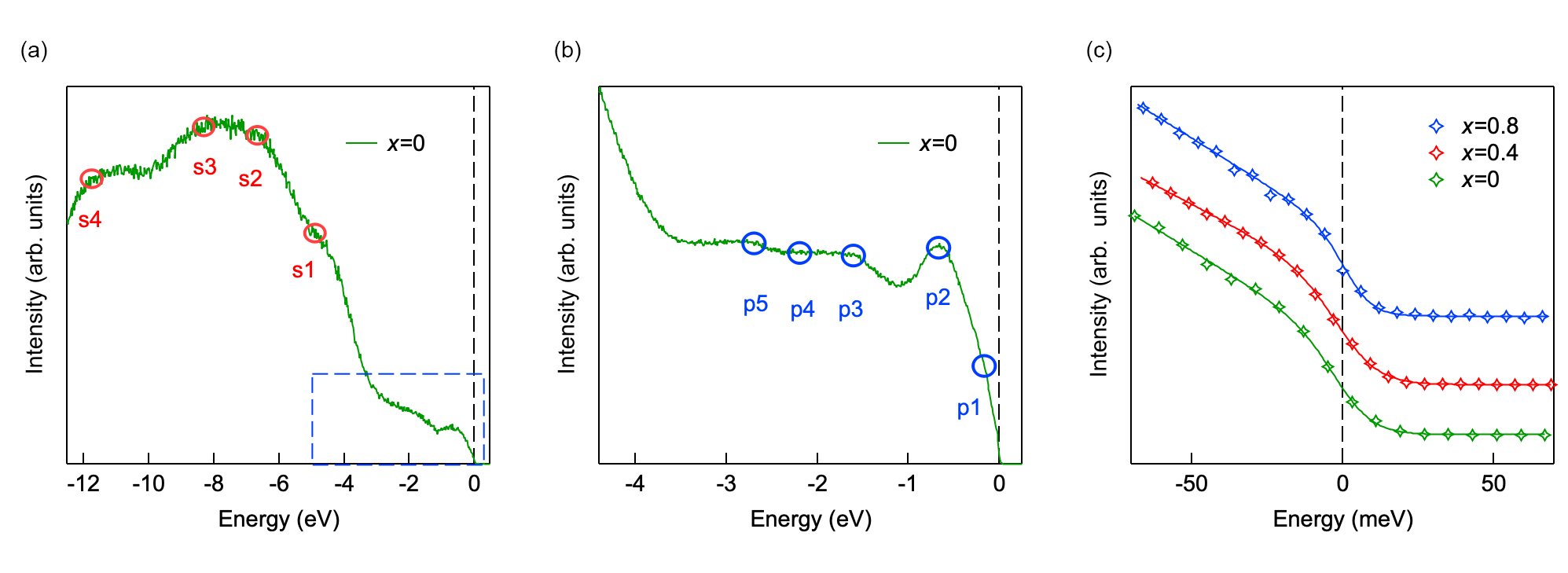}
\caption{(a) SDOS derived from momentum-integrated ARPES data collected using HeII$\alpha$ radiation ($h\nu\sim40.8$eV) at 300K. The valence band peaks, labeled s1, s2, s3 and s4, relate to s orbitals of Ge, Sb and Te. The energy range marked by the blue dashed rectangle corresponds to the locations of the valence band peaks derived from p orbitals of Ge, Sb and Te. All these peaks in (a) are not clearly resolved. (b) SDOS at 20K collected with HeI$\alpha$ radiation ($h\nu\sim21.2$eV) clearly display the these peaks, labeled as p1, p2, p3, p4 and p5, associated with p orbitals of Ge, Sb and Te. (c) SDOS at 20K close to the chemical potential has a Fermi step for for $x=0,0.4,0.8$. The data for each $x$ can be fit by a resolution-broadened Fermi function, shown by solid lines, indicating that the samples are in the metallic regime. Note that the data from various $x$ values have been normalized and offset for visual clarity.}
\end{figure}

\begin{figure}
\includegraphics{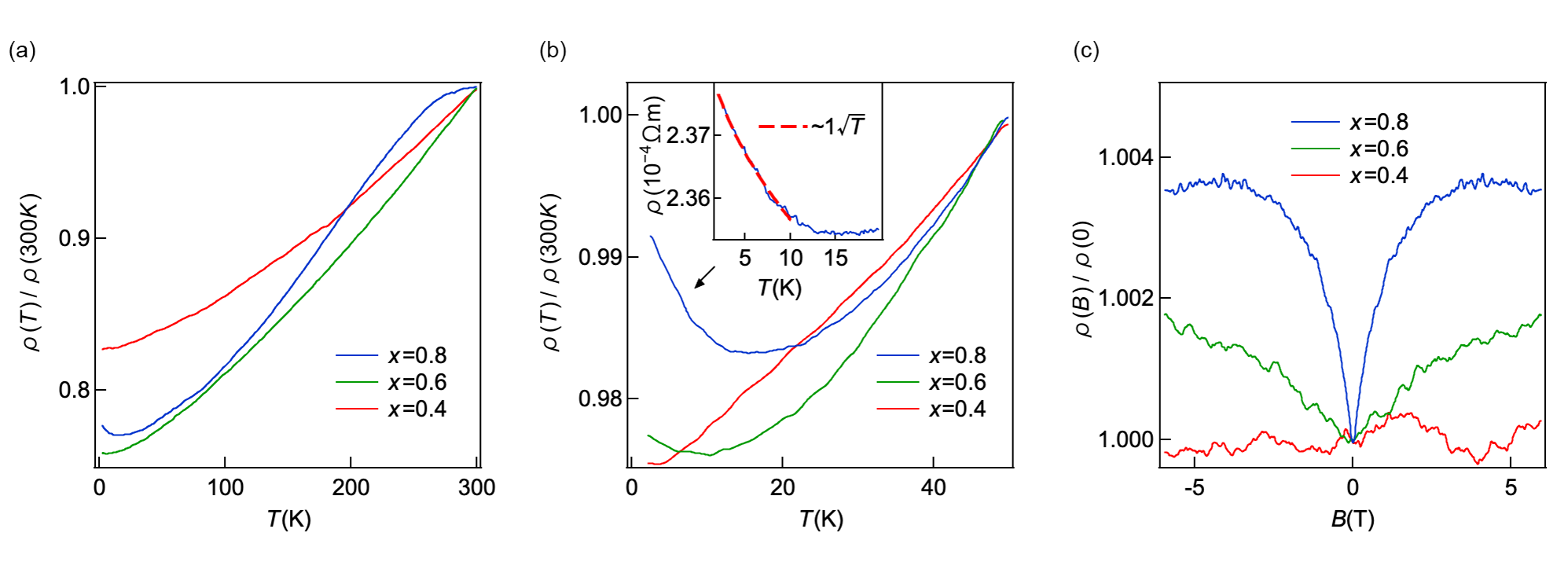}
\caption{Temperature dependence of the normalized resistivity from 2K up to (a) 300K and (b) 50K for $x=0.4,0.6,0.8$. The onset of a resistance upturn below $T\sim20$K appears as doping is increased $x\geq 0.6$. (c) Normalized magnetoresistivity at 2K up to $\pm$6T, displaying a resistance minimum around $B=0$ for $x\geq 0.6$. Both of these features grow as $x$ is increased.}
\end{figure}

\begin{figure}
\includegraphics{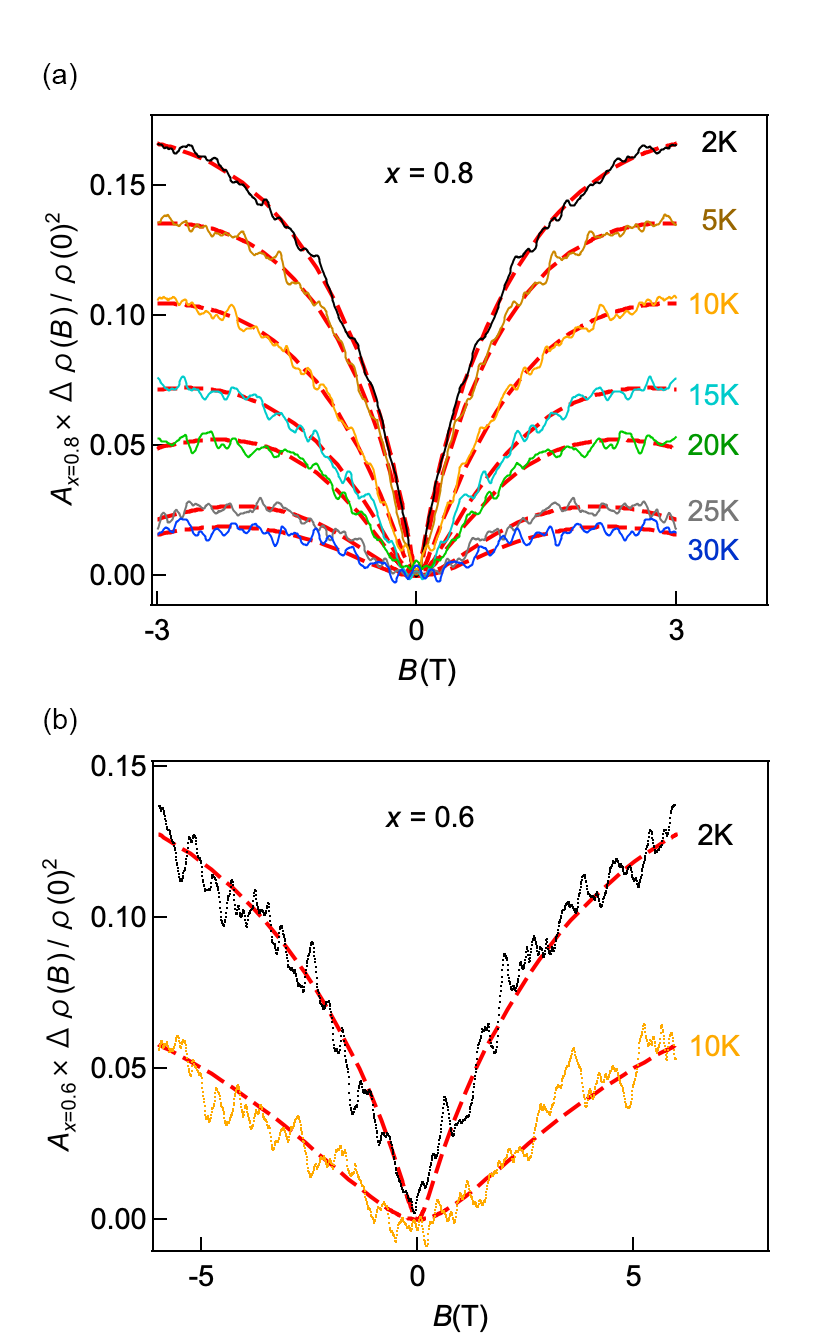}
\caption{Least-squares fitting of the magnetoresistance to the Fukuyama-Hoshino model for (a) $x=0.8$ and (b) $x=0.6$. Dashed lines represent fitting results, which provide an estimate for the phase coherence length $l_\phi$.}
\end{figure}

\begin{figure}
\includegraphics{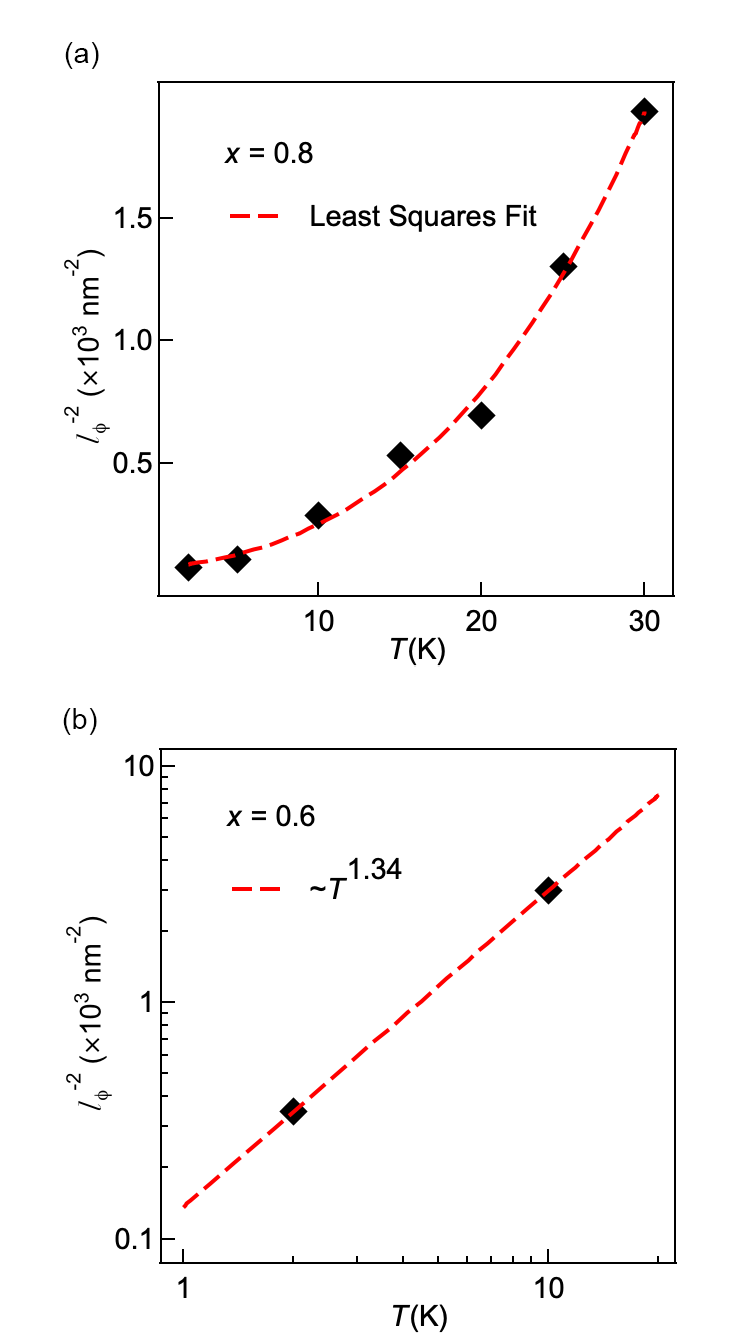}
\caption{Temperature dependence of the phase coherence length for (a) $x=0.8$ and (b) $x=0.6$, extracted from the Fukuyama-Hoshino model (see main text). In (a), the dashed line represents a fit to Equation 3, from which we find that $C_{ee}$ is nearly two orders of magnitude larger than $C_{eph}$. In (b), the dashed line represents a fit to a power law $l^{-2}_{\phi}(T)\sim~T^{1.34}$, which is close to the expected power law when electron-electron Coulomb interactions dominate dephasing.}
\end{figure}

The single-particle density of states (SDOS) data we obtained from ARPES for various $x$ values are displayed in Figure 1. Given that previous SDOS studies on GST have been conducted on thin films, it is important to compare the overall SDOS features from thin film samples with those from our bulk samples. To this end,  in Figure 1a and Figure 1b, we show the SDOS data collected from $x=0$ using HeII$\alpha$ radiation ($h\nu\sim 40.8$eV) and HeI$\alpha$ radiation ($h\nu\sim 21.2$eV), respectively. Figure 1a, which shows data taken at 300K, clearly shows the high binding energy valence band SDOS peaks (labeled as s1, s2, s3 and s4) associated with s orbitals of Ge, Sb and Te in the energy range between 4.5eV and 12.5eV. The SDOS peaks corresponding to p orbitals of Ge, Sb and Te appear between 0 and 4.5eV, and are not completely resolved in Figure 1a.  All these peaks (labeled as p1, p2, p3, p4 and p5) are, however, better resolved in Figure 1b, taken at lower energy $h\nu\sim 21.2$eV and at a lower temperature of 20K. These features are overall consistent with what has been observed in both ultraviolet photoelectron spectroscopy (UPS) and X-ray photoemission data from thin films samples of similar compounds\cite{richter_hard_2014,kellner_mapping_2018,pauly_evidence_2013,sarkar_origin_2018,klein_changes_2008,kim_electronic_2007}. There is, however, a notable difference between the SDOS data from our bulk GST samples and those from the thin film samples---unlike in the thin film samples, there is no energy gap in the SDOS of our bulk GST sample at the chemical potential. This is also the case for bulk GSST samples ($x>0$) at each measured $x$. This can be clearly seen in Figure 1c, where we present higher resolution data at 20K within a very narrow energy window around the chemical potential $\mu$, collected with $h\nu\sim 21.2$eV, for $x=0$, $x=0.4$, and $x=0.8$. The Fermi step, a characteristic spectroscopic feature of a metallic sample, can be inferred from the fact that the data for each measured $x$ is well approximated by a resolution-broadened Fermi function (shown by solid lines in Figure 1c). To the best of our knowledge, this is the first reported spectroscopic signature of a truly metal-like behavior in either GST or GSST samples. This in turn supports the degenerately doped semiconductor picture for this material in its bulk form. This is in  agreement with the transport data, discussed below, which suggests that each of the measured samples are in the metallic regime. 

Transport measurements were performed at temperatures from 2K to 300K and under perpendicular magnetic fields $B$ up to $B=6$T. Samples were current biased with a current of 1mA and the resistance was extracted using a standard four probe, low frequency lock-in technique. In Figure 2a-b, we plot the temperature dependence of the normalized resistivity at zero magnetic field (resistivity in raw units can be found in the Supplemental Material). It can be seen that the resistivity $\rho$ varies significantly as the Se doping $x$ is increased, with a low $T$ resistance upturn appearing below $\sim20$K as $x$ is increased above 0.6.

Such a low $T$ upturn may indicate the presence of a Kondo-like effect \cite{kouwenhoven_revival_2001}, where magnetic impurities enhance disorder scattering as temperature is reduced. However, as the Se dopants are nonmagnetic, we instead model this behavior by considering the theory of conduction in disordered 3D metals \cite{lee_disordered_1985}, in which the conductivity $\sigma(T)=1/\rho(T)$ can be written as:

\begin{equation}
\sigma(T) = \sigma_0 + K\sqrt{T},
\end{equation}

\noindent where $\sigma_0$ is the residual (Drude) conductivity and $K$ is a constant (here, we assume $\sigma_0$ and $K$ depend on $x$). The second term, involving a square-root power law in $T$, originates from electron-electron Coulomb interaction (EEI) effects.

In our case, a least-squares fitting to a power law $\rho(T)=C_1+C_2T^{\gamma}$, where $C_1$ and $C_2$ are constants, for $x=0.8$ in the range 2K$<T<$10K gives an exponent $\gamma= -0.40$, very close to the expected value of $-1/2$ for EEI effects (see Equation 1). Indeed, as shown in Figure 2b (inset) a square-root power law reproduces the observed resistance trend very well. This is consistent with prior studies on thin films of pristine GST, where EEI effects also led to a low $T$ resistance upturn \cite{breznay_weak_2012}. We therefore attribute this trend to the onset of weak/moderate electronic interactions. 

Despite this observation, the small deviation from a square-root power law shown above suggests that other mechanisms could also contribute to the measured resistance. To gain a greater understanding of such mechanisms, in Figure 2c we plot the magnetoresistance (MR). At a smaller doping of $x=0.4$, the MR is flat up to $B=\pm6T$, while for larger doping, a minimum in resistance appears around $B=0$. This minimum becomes deeper with increasing doping.

To understand this behavior, we consider disorder-induced quantum interference (QI) effects. In many metallic systems, these effects lead  to weak localization, which occurs because disorder scattering creates a nonzero probability for an electron to propagate along a closed loop. In the presence of time-reversal symmetry, the quantum mechanical amplitude gained by the electron traversing the loop and that of its time-reversed partner interfere constructively (so long as quantum phase coherence is preserved), leading to an increased probability for the electron to localize. The breaking of time-reversal symmetry by a small magnetic field generates a resistance peak in the MR around $B=0$. On the other hand, if spin-orbit coupling (SOC) is present, this produces an additional $\pi$ phase shift in the electron wavefunction as it goes around a closed path, due to a SOC-induced spin flip, leading to weak anti-localization that manifests as a dip in the MR at  $B=0$. Such strong SOC is likely introduced by the Se substitution of Te in our samples, accounting for the positive MR trend we observe.

To analyze our data more quantitatively, we note that in 3D the change in magnetoresistance due to QI with SOC is described by the Fukuyama-Hoshino model \cite{fukuyama_effect_1981}:

\begin{equation}
    \left( \frac{\delta \rho}{\rho^2} \right) = \frac{e^2}{2\pi^2\hbar} \sqrt{\frac{eB}{\hbar}} \left\{ \frac{1}{2}f_3\left(\frac{B}{B_{\phi}}\right) - \frac{3}{2}f_3\left(\frac{B}{B_2}\right) \right\}.
\end{equation}

\noindent Here,

\begin{equation*}
    B_{\phi} = B\textsubscript{i} + 2B\textsubscript{s}, \hspace{2mm} B_2 = B_i + \frac{2}{3}B\textsubscript{s} + \frac{4}{3}B\textsubscript{so}
\end{equation*}
\begin{equation*}
    B_\textsubscript{i} = \frac{\hbar}{4eD\tau_\textsubscript{i}}, 
     B_\textsubscript{s} = \frac{\hbar}{4eD\tau_\textsubscript{s}}, 
      B_\textsubscript{so} = \frac{\hbar}{4eD\tau_\textsubscript{so}},
\end{equation*}

\noindent where $D$ is the electronic diffusivity and $\tau_j$ for $j =$ i, so, and s refers to the inelastic, spin-orbit, and magnetic spin-flip scattering times, respectively. $f_3$ is the Kawabata function, given by 

\begin{equation*}
    f_3(z) = \sum^{\infty}_{n=0} \left\{ 2 \left( n+1+\frac{1}{z} \right)^{1/2} -2\left(n+\frac{1}{z}\right)^{1/2} \left(n+\frac{1}{2}+\frac{1}{z}\right)^{-1/2} \right\} .
\end{equation*}

An important parameter in this model is the phase coherence length, $l_\phi=D\tau_\phi$, which provides an estimate for how far electrons can travel before losing their quantum phase coherence ($\tau_\phi$ is the phase relaxation time, defined by the relation $B_{\phi} = \hbar/(4eD\tau_{\phi})$, see Equation 2). To estimate this length, we perform a least-squares fitting of the MR to Equation 2. Here, we set $B_s=0$ (since this material is nonmagnetic), and also include a prefactor $A_x$ to account for the sample dimensions. The fitting results are shown in Figure 3 at various temperatures.

For $x=0.8$ and $x=0.6$ the phase coherence length is determined to be 72nm and 54nm, respectively, at $T=2$K. In both cases, $l_\phi$ is much smaller than the sample thickness ($\sim$5mm), affirming the 3D nature of our samples down to cryogenic temperatures. 

Next, we turn to a determination of the particular quantum processes limiting the phase coherence length. Typically, inelastic processes such as electron-electron or electron-phonon scattering contribute to dephasing \cite{lin_recent_2002}. Here, we study the temperature dependence of $l_\phi$ to help disentangle the two. 

In 3D, $l_\phi$ is expected to follow a power law in temperature, given by \cite{lin_recent_2002}:

\begin{equation}
    l_{\phi}^{-2}(T) = l_{\phi}^{-2}(T=0) + C_{ee}T^{p_{ee}} + C_{eph}T^{p_{eph}},
\end{equation}

\noindent where $p_{ee}$ and $p_{eph}$ are exponents corresponding to the dephasing processes due to electron-electron and electron-phonon scattering, respectively. $p_{ee}= 3/2$, while $p_{eph}$ is 2 or 3 for inelastic scattering by transverse or longitudinal phonons, respectively. From Figure 3, we extract the temperature dependence of $l_\phi$ and plot the results, shown in Figure 4. For $x= 0.8$, we obtain the best fit to Equation 3 by assuming that longitudinal phonons dominate over transverse ones, $p_{eph}= 3$, allowing us to obtain prefactors $C_{ee}= 4.15 \times 10^{12}$ and $C_{eph}= 4.37 \times 10^{10}$. This fitting is shown as red dashed lines in Figure 4a. Interestingly, $C_{ee}$ is nearly two orders of magnitude larger than $C_{eph}$, suggesting that the dephasing is primarily due to EEI effects in this material.

\noindent For $x=0.6$, the MR minimum is significantly weaker and vanishes at much smaller $T$, prohibiting a complete fitting to Equation 3 for our limited data set. However, fitting to a power law $l^{-2}_{\phi}(T)= C_1+C_2T^{\beta}$ yields $\beta=1.34$. This value of $\beta$ is very close to the expected EEI term $p_{ee} = 3/2$, again suggesting that dephasing is primarily due to EEI effects at even smaller dopings $x=0.6$. Coupled with the temperature dependence (Figure 2a-b), we conclude that EEI effects play a noticeable role in the low-$T$ physics of this material.

EEI effects have been shown to lead to a myriad of electronic phases, including high temperature superconductivity \cite{bednorz_possible_1986}, Wigner crystallization \cite{wigner_interaction_1934}, and charge-density wave states \cite{wang_competing_2013}. In most cases, these unique phases lead to emergent electronic behavior that could broaden the potential set of technological applications for these materials. In our experiment, the strength of electronic interactions is readily tuned by the Se dopants, as is evident by the magnetoresistance trend as well as the strength of the low $T$ upturn. While no obvious correlated phases are studied in this work, the tunability of EEI effects with Se dopants nevertheless demonstrate that GSST can be an excellent material to study the influence of electronic interactions on a prototypical 3D PCM.

Finally, we consider the role of Anderson localization in our experiment. To this end, the Ioffe-Regel criterion is often used \cite{overhof_electronic_1989, siegrist_disorder-induced_2011}. This condition states that when the mean free path of electrons $l$ is reduced via disorder scattering to be smaller than the Fermi wavelength $2\pi/k_F$ ($k_F$ the Fermi wavevector) the electrons become localized, and a MIT occurs. In most materials, this generally occurs around $k_Fl \sim 1$. Indeed, such a disorder-induced MIT has been seen in many materials \cite{mott_metal-insulator_1985,dai_electrical_1992}, including GST \cite{siegrist_disorder-induced_2011}. Using the relation

\begin{equation}
k_fl=\frac{\sigma\hbar}{e^2}\frac{k_F^2}{n},
\end{equation}

\noindent and using the extracted Hall density $n$ from our samples (supplementary material), we determine $k_F l=8$ and $k_F l =3$ for $x=0.6$ and $x=0.8$, respectively, at $T= 2$K. This value of $k_F l$ indicates the sample is on the metallic side of the MIT but very close to the critical point, consistent with the observed weak anti-localization behavior. These results are moreover consistent with the ARPES observation of gapless SDOS for $0\le x \le 0.8$.

To summarize, both ARPES and electronic transport measurements suggest metallic behavior in bulk GSST at all measured $x$ in the range $0\le x \le 0.8$. Transport data further presents evidence for QI effects with SOC that is likely introduced by the Se dopants, as well as evidence for EEI effects, when $x\geq 0.6$. These effects grow with Se doping. Interestingly, the Ioffe-Regel condition places this material very close to the critical point, implying strong disorder scattering together with the observed Coulomb interaction effects. To our knowledge, there is no complete theoretical framework which fully captures the interplay between strong disorder scattering and electron-electron interactions. Further experimental and theoretical work will therefore be necessary to fully understand the resulting electronic physics that can emerge in this novel phase change material.

We thank Nandini Trivedi for helpful discussions. Devices are fabricated using the nanofabrication facility that is supported by NSF Materials Research Science and Engineering Center Grant DMR-2011876. UC was partially supported by the National Science Foundation under Grant No. DMR-1454304. ZX and DL were supported by the Department of Energy, Grant number DE-FG02-01ER45927.

Detailed experimental methods, Hall density data, and magnetic field dependence of the $R(T)$ curves is available in the supplementary material (PDF).

This article has been submitted to Applied Physics Letters. After it is published, it will be found at https://pubs.aip.org/aip/apl.

\bibliographystyle{aipnum4-1}
\bibliography{aiptemplate.bib}

\end{document}